\documentclass[apj]{emulateapj} \setkeys{Gin}{width=0.48\textwidth}
\usepackage{graphicx}

\newcommand{\hmpc}{\ifmmode{h^{-1}\,\hbox{Mpc}}\else{$h^{-1}$\thinspace Mpc}\fi}

\newcommand{\kms}{\ifmmode{\,\hbox{km\,s}^{-1}}\else {\rm\,km\,s$^{-1}$}\fi}
\newcommand{\msun}{{\rm\,M_\odot}}

\begin{document}
\title{Globular Clusters in a Cosmological N-body Simulation}
\shorttitle{Cosmological Globular Cluster Simulations}
\shortauthors{Carlberg}
\author{Raymond G. Carlberg}
\affil{Department of Astronomy \& Astrophysics, University of Toronto, Toronto, ON M5S 3H4, Canada} 
\email{carlberg@astro.utoronto.ca }

\begin{abstract}
Stellar dynamical model globular clusters are introduced into
reconstituted versions of the dark matter halos of the Via-Lactea II (VL-2) simulation 
to follow the star cluster tidal mass loss and stellar stream formation.
The clusters initially evolve within their local sub-galactic halo, later being accreted into the main halo.
Stars are continually removed from the clusters, but those that emerged in the sub-galactic halos are
dispersed in a wide stream when accreted into the main halo.  
Thin tidal streams that survive to the present can begin to form
 once a cluster is in the main halo.
A higher redshift start places the star clusters in  denser halos 
where they are subject to stronger tides leading to higher average mass loss rates.
A z=3 start leads to  a rich set of star streams with  nearly all within 100 kpc having
a remnant progenitor star cluster in the stream.
In contrast, in a z=8 start, all star
clusters that are accreted onto the main halo are completely dissolved.  
These results are compared to the available data on Milky-Way streams,
where the majority of streams do not have clearly associated globular clusters.
which, if generally true, suggests that there were at least twice as many
 massive globular clusters at high redshift.
\end{abstract}


\section{INTRODUCTION}
\nobreak
Present day  halo globular clusters are 
a tracer population in the dark matter of the Milky Way, providing important insights
into the assembly history of the galaxy and 
some of the first observational evidence for galactic dark matter \citep{HS:78}.
\citet{SZ:78} proposed a basic conceptual picture in which star clusters formed 
in  ``transient protogalactic fragments" which accrete onto the developing galaxy 
to form the stellar and dark matter
halo of the Milky Way galaxy. This picture is now elaborated within general models of hierarchical galaxy
formation within the Cold Dark Matter (CDM) cosmology.

Globular clusters evolve through the gravitational interactions of their constituent stars and the influence
of external gravitational fields.
Encounters between individual stars, binaries and more complex bound units, systematically increases the binding 
energy of the central region causing the outer regions of the star cluster to expand.
The external gravitational field, which can often be approximated with a tidal tensor, 
set a distance beyond which cluster stars are no longer
bound to the cluster. And, time varying tidal fields, from either other massive sub-halos or peri-galacticon passage,
pump energy into cluster stars. 
The internal two-body relaxation time increases with increasing mass \citep{Spitzer:87} for the same
orbit within a galaxy. 
Consequently, lower mass clusters lose mass more quickly, systematically
depleting the lower mass clusters so that over the
lifetime of the galaxy a wide range of initial mass functions come
to resemble the nearly universally observed log-normal mass distribution
 \citep{FR:77,GO:97,FZ:01}. 

The stars that become unbound from their parent cluster are pulled away in tidal
streams that are offset in angular momentum so lead and trail the cluster in its orbit. 
In principle a stream provides a complete history of the globular 
cluster and the properties of the potential along its orbit. 
The density of stars along a tidal stream provides an
empirical estimate of the mass loss from its parent cluster, which
in principle could be used to reconstruct the cluster, including completely dissolved clusters. 
The simulations here will shed light on some of the difficulties that will need to be overcome to allow even
a statistical reconstruction of the history of interest.

The thin components of streams 
are sensitive indicators of the low mass sub-halos that 
the LCDM power spectrum predicts should be orbiting in the halos of the galaxy 
\citep{Klypin:99,Moore:99,YJH:11,Carlberg:12,Bovy:17}. Current methods of analysis have concentrated on
the relative simple streams that develop in a static potential, which may require modification to
take into account
the time variations of the gravitational fields in which the stream develops.

High resolution gas dynamical galaxy formation simulations 
are beginning to explore formation of dense star clusters within a galaxy
\citep{KG:05,BDE:08,Kimm:16,RAG:17,Li:17,Kim:17}. The current spatial resolution of the
resulting clusters is not yet adequate to follow the dynamical evolution of individual
stars in the clusters. The study of star clusters remains a challenging numerical problem, largely
addressed through the \citet{Aarseth:99} program of gravitational n-body 
codes without particle softening. The current version, Nbody6 and its specialized
variants, is a highly realistic cluster code, which allows external fields to be described with
an externally supplied time varying tidal tensor.

The numbers and masses of globular clusters at high redshift are of interest for the reionization of the
universe \citep{Bouwens:17,BK:17} via the the ionizing radiation from their massive stars present
at early times, and, the creation of binary black holes which inspiral to merge and emit gravitational waves
at low redshift. 
The low metal abundance of globular cluster stars reduces the 
stellar winds of the high mass stars which can lead to stellar remnant black holes in the $30\msun$ range.
The black holes sink via gravitational interactions into the central regions where processes of 
binary exchange and binary gravitational hardening over a Gyr or so  can put together 
massive binary black holes (BBHs) that will inspiral over a Hubble time to merge \citep{PZM:00}. 
BBH merger rates predicted from the current
population of globular clusters generally come up somewhat low \citep{Rodriguez:15,Rodriguez:16,Chatterjee:17,Askar:17} relative
to the currently measured LIGO event rates \citep{LIGO2}. If
the high redshift population of globular clusters were significantly larger than
the low redshift numbers on which current predictions are based. 
It is interesting that both BBHs and reionization may be connected through the high mass 
stars in high redshift globular clusters.

To better understand the history of globular clusters and their tidal streams this paper follows 
a population of globular clusters formed at high redshift
in the sub-galactic dark matter halos that merge over time to create a galactic halo.
The evolution of a pre-galactic dark matter distribution drawn from
a larger cosmological simulation to form a galaxy's dark matter halo 
is a well understood n-body simulation problem. Here such
a simulation is augmented through the insertion of 
dense star clusters in the central regions of the pre-galactic sub-halos. The added star clusters contribute so little 
mass to the overall simulation, less than 0.01\% here,  that the dark matter history is barely affected.
The simulation is conceptually started at a time after the massive stars in the clusters 
have evolved and dispersed any surrounding gas so that hydrodynamical effects no longer
play a role. The star particles use a gravitational softening to speed up the calculation, requiring
that the expected level of two-body relaxation between stars in the clusters be added in 
with a Monte Carlo velocity perturbation scheme.

Questions of interest here are to what degree the visible tidal streams be used to reconstruct the
initial dense stellar cluster distribution, and,  what is the character of 
dark matter sub-halo induced gaps in these streams. 
Both of these questions can be answered without a high precision
cluster mass loss model, although the rate of mass loss must be realistic.
The next section lays out the details of the simulations with the starting conditions for the
dark matter, the construction of the individual star clusters and where they
are inserted into the dark matter distribution. The Monte Carlo scheme to 
appropriately add two-body relaxation back into the softened star clusters
along with high accuracy calibration runs is described in Section 2.2.
Section 3 describes the evolution of the cluster masses and how the
stars pulled away from the clusters are distributed in streams.
 
\begin{figure}
\begin{center}
\includegraphics[angle=0,scale=0.8,trim=190 190 190 190,clip=true]{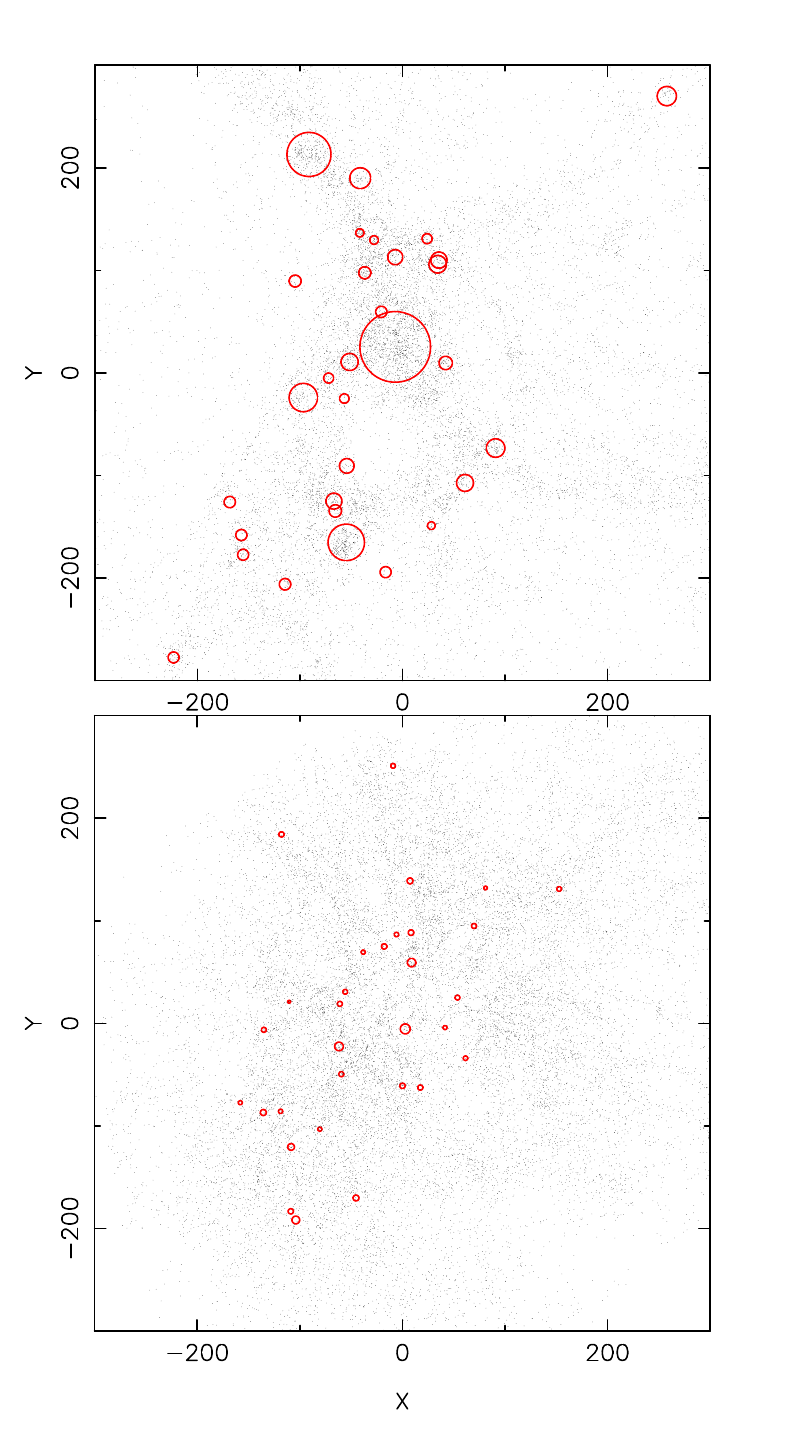}
\end{center}
\caption{The VL-2 halos at z=3, 2.078 Gyr (top) and z=8, 0.70 Gyr (bottom).
The red circles are the halos into which the star clusters will be inserted, with the circle radius
being equal to the $R_m$ of the halo.  The remaining halos
are shown as dots at the halo center. No particles are shown.
}
\label{fig_xy0}
\end{figure}

\begin{figure}
\begin{center}
\includegraphics[angle=0,scale=0.8,trim=190 190 190 190,clip=true]{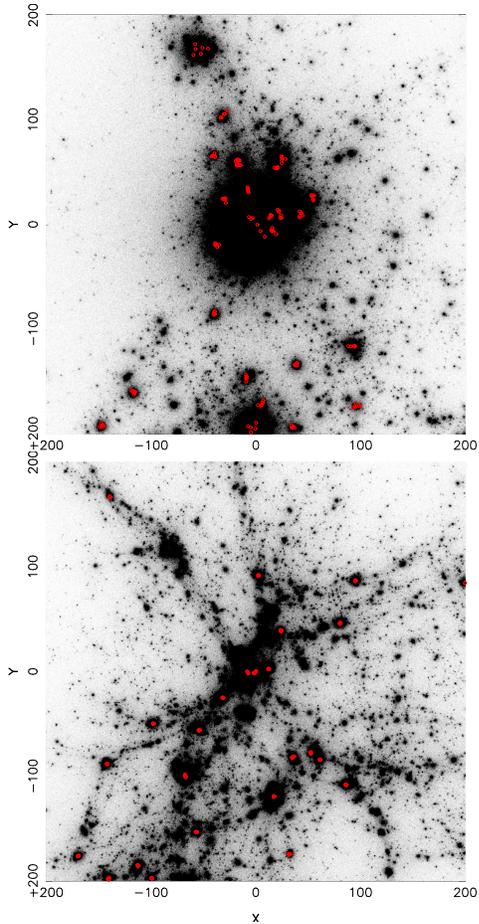}
\end{center}
\caption{The reconstituted dark matter particle distribution 
0.5~Gyr after starting at z=3  (top) and z=8 (bottom panel). The red circles are the locations
of the centers of the model star clusters. }
\label{fig_xy50}
\end{figure}

\section{Globular Clusters in a Cosmological Simulation}

The formation processes that lead to the low metallicity halo population of globular clusters
of the Milky Way and all other galaxies are not yet well understood. 
At the high redshifts corresponding to their stellar ages
the most common host dark matter halos will have peak circular velocities in roughly the 10-30\kms\ range,
sufficient to retain gas as reionization occurs, but the gas will largely be driven out as a result energy and momentum
input from the more massive stars that they form, leaving a largely dark matter dominated system 
containing a low metallicity stellar population \citep{DS:86}.
Accordingly, it is a reasonable step to focus simulations on the dynamical processes of star clusters in dark matter halos,
temporarily setting aside the complications of gas and the stellar structure of the host halos. 
The star clusters are conceptually placed in the dark halos 
after their most massive stars have gone through their short and dramatic evolution in 
the first few hundred million years of the star cluster's evolution, after which
stellar evolution produces smaller and slower changes in the mass of stars.
The initial conditions then become fairly straightforward, requiring a dark matter distribution at high redshift which evolves into 
a good approximation of the Milky Way's dark halo. An appropriate subset of the numerous low mass dark halos present
at high redshift is then populated with dense star clusters. Subsequent
gravitational evolution creates a Milky Way like dark halo containing the remnant star clusters
and their tidal tails.
Most of the halo clusters orbit at sufficiently large radius that the galactic disk and bulge do not substantially disturb the clusters.

The simulations contain a mixture of star particles and dark matter particles. The star particles
begin in star clusters that are inserted within the top thirty or so dark matter halos identified
in the VL-2 simulation \citep{VL2}.
The  resulting mixture of particles is followed using the gravity code Gadget2 \citep{Gadget2}, augmented 
to provide the two-body relaxation between stars that drives cluster evolution.  

\subsection{The Cosmological N-body Halos}

The initial conditions for the simulations use the snapshots of the Via Lactea II simulation (VL-2) \citep{VL2}.
A set of 20048 halos, with masses above about $2\times 10^5\msun$,
 were identified within VL-2 at  redshift zero and then traced back in time 
to give their size, mass, position and velocity.  
The VL-2 data files for these progenitor sub-halos are the basis of
the initial conditions here. Two
snapshots, near redshifts 3 and 8, are selected as start times for the insertion of
dense star clusters. The lower redshift is within epoch of peak star formation for galaxies 
and the higher redshift is within the epoch of reionization, which roughly spans 
the likely range of times over which the old halo clusters are formed.
A continuous distribution of globular cluster formation 
times could be recreated by summing over appropriately weighted starting times, 
since the star clusters are sufficiently light
that they have essentially no effect on the dark matter halos. 
Only those sub-halos above a minimum sub-halo mass, in the range of a few $10^9\msun$,
are normally populated with star clusters,  leaving the rest empty.
To create the distribution of
particle positions and velocities the individual dark matter halos 
are approximated as \citet{Hernquist:90} spheres.
Each halo is characterized with a maximum circular
velocity, $V_m$, which occurs at radius $R_m$, together giving
a  Hernquist mass  $M_H = 4 G^{-1}V_m^2 R_m$. 
At time 0.70 Gyr, redshift 7.8,
halos with $M_H \ge 0.1\times 10^{10}\msun$ are selected,  which identifies
31 halos, and at time 2.08 Gyr, redshift 3.24, $M_H\ge 0.5\times 10^{10}\msun$ which selects
33 halos. The selected halos at the two times are shown in Figure~\ref{fig_xy0}.

\begin{figure}
\begin{center}
\includegraphics[angle=0,scale=0.8,trim=180 190 190 190,clip=true]{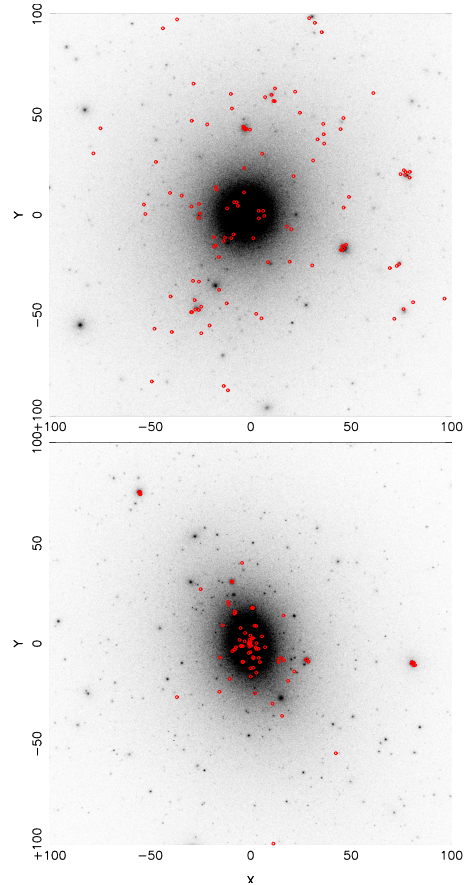}
\end{center}
\caption{The dark matter at the end point for the  z=3  (top) and z=8 (bottom) start systems.
The red circles are the centers of the globular clusters. The boxes are 60 kpc on an edge.
}
\label{fig_xydend}
\end{figure}

The total mass of the 20048 halos in the simulation is  $2.54\times 10^{12}\msun$.
The total mass in the halos at early times  is 
50\%  and 5\% of  the final total mass 
at z=3 and z=8, respectively. To allow for the mass that is not in the Hernquist sphere halos, 
the additional mass is added as an extended, low density, sphere of particles centered on the 20048 primary halos.
This starting setup is sufficiently similar to the VL-2 simulations that its gravitational evolution
 to the dominant, sub-halo rich, Milky Way-like final halo. 

The detailed procedure to create the particle distribution
first uses the \citet{Hernquist:90} mass distribution with radius to give
the particle positions. The analytic 
distribution function as a function of energy 
 for the Hernquist  mass distribution gives the velocity distribution at each radius from
which an equilibrium particle distribution for each halo is created.
The extra mass required for the total mass is distributed in proportion to the primary halo mass and placed in a 
second Hernquist sphere given a scale radius 8 times the radius of
the central equilibrium distribution, with velocities reduced to
80\% of equilibrium in the low density region to ensure that they are accreted
onto the central halo over time. The secondary extended halo is low enough density 
that it does not substantially disrupt the primary halo. 
The reduced random
velocities allow an accretion flow onto the central halo to develop, but is
sufficient to suppress a  coherent rapid collapse and local gravitational instability.
The dark matter distributions of the two simulations are shown 1 Gyr into the simulation in 
Figure~\ref{fig_xy50}. The z=8 start still has strong cosmic web features, whereas
the z=3 start with its dominant halo is in the regime of accretion of satellites.
 

\subsection{Simulating Dense Star Clusters}

The internal evolution of star clusters sufficiently dense
to survive for a Hubble time
remains a vigorous research area. A primary goal of this study is to accurately follow the stars
that leave the cluster in the tidal streams. 
A secondary goal is to study the mass evolution of clusters as a function of their cosmological history.
All  the internal structural details of the star clusters are not needed, but the clusters need 
to be sufficiently realistic to lose mass at approximately the correct rate. A simplified physical model
of cluster star-star encounters is developed with its parameters calibrated using a high accuracy
gravitational n-body model.

\subsubsection{Star Cluster Internal Evolution Model}

Star clusters evaporate as a result of gravitational encounters of stars
\citep{Henon:61,Spitzer:87}.  Following the evolution of a single star cluster
in full detail remains a substantial computational challenge \citep{Aarseth:99}.
\citet{Carlberg:17} presented a model in which a
softened gravity cluster is augmented with a simplified single-zone heating model
to approximate the relaxation which drives evaporation. 
The model calculates  velocity from the two-body relaxation time,  which is then randomly
added to the stars within some fraction of the tidal radius of the model star cluster.
The  calculation of the required velocity increment is
done at intervals of $\delta t$.
The velocity increment for a cluster of characteristic mass $M_c$ is calculated using the RMS 
velocity diffusion coefficient \citep{BT:08},
\begin{equation}
\delta v =     {\sigma_c \over C_h}  \left({\delta t\over 4.9\ {\rm Myr} } \right)^{1/2} 
 \left({M_c\over 10^{5}\msun}\right) ^{1/4} 
\left({r_v\over 10\  {\rm pc}}\right)^{-5/4},
\label{eq_dvheat}
\end{equation}
where $\delta v$ is in units of \kms,  $\sigma_c$ is the
cluster velocity dispersion, and $r_v$ is the virial radius, calculated with softening included, then
removed in quadrature from the result. The weak explicit dependence 
on cluster mass, $M_c$, is useful to retain to ensure that the heating remains reasonable as the 
clusters near complete evaporation.

The cluster 
mass,  $M_c$ and velocity dispersion are measured within some fiducial radius which is set to be some multiple of the
virial radius, usually set at $6r_v$. The virial radius is preferred over the half mass radius on the basis that it is simpler
to calculate and is a stable quantity with a weak dependence on the maximum radius. The clusters  here
have $r_v$ values about 2/3 of $r_h$. 
The kinematic properties are calculated for each cluster each time the velocity increments 
need to be added to the star particles. The random velocities are added to all star
particles within a radial range of  typically 0.5-1.5$r_v$, with no disturbance of more distance particles.
A dummy particle with nearly zero mass is used to mark the center, which speeds
up the calculation.  At 
each heating step the dummy's location is updated to the current center of mass and momentum.
The $\delta v$ calculated from Equation~\ref{eq_dvheat}
 is generates a Gaussian distribution of velocity offsets in each velocity direction.

\subsubsection{N-body Cluster Calibration}


Nbody6  direct n-body simulations \citep{Aarseth:99} are used to calibrate 
the model heating parameters. 
The Nbody6 code is used to generate a Plummer model with 20,000 particles, all of equal mass of 0.7543267 $\msun$. 
The cluster of particles is placed in a logarithmic potential with a circular velocity of 240\kms 
at 30 kpc from the center and given an initial tangential velocity of 120\kms. 
The circular velocity is close to the value of the final dark matter halo of the simulation 
and the orbit is fairly typical for both real world star clusters that produce streams and 
many of those that the simulations create.

\begin{figure}
\begin{center}
\includegraphics[angle=-90,scale=0.7,trim=20 40 20 20,clip=true]{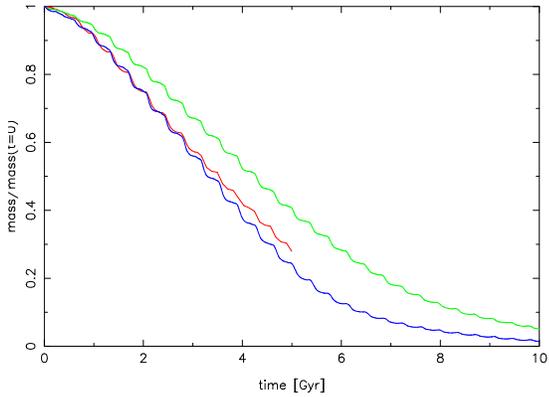}
\end{center}
\caption{An N=20,000 Plummer sphere of equal mass stars having a mass $1.51 \times 10^4\msun$ created evolved with
Nbody6 (red line) and the heated cluster code used here
with the heating parameter set at 25 (the adopted value, blue) and 30 (green).
}
\label{fig_mtplum}
\end{figure}

\begin{figure}
\begin{center}
\includegraphics[angle=-90,scale=0.7,trim=20 40 20 20,clip=true]{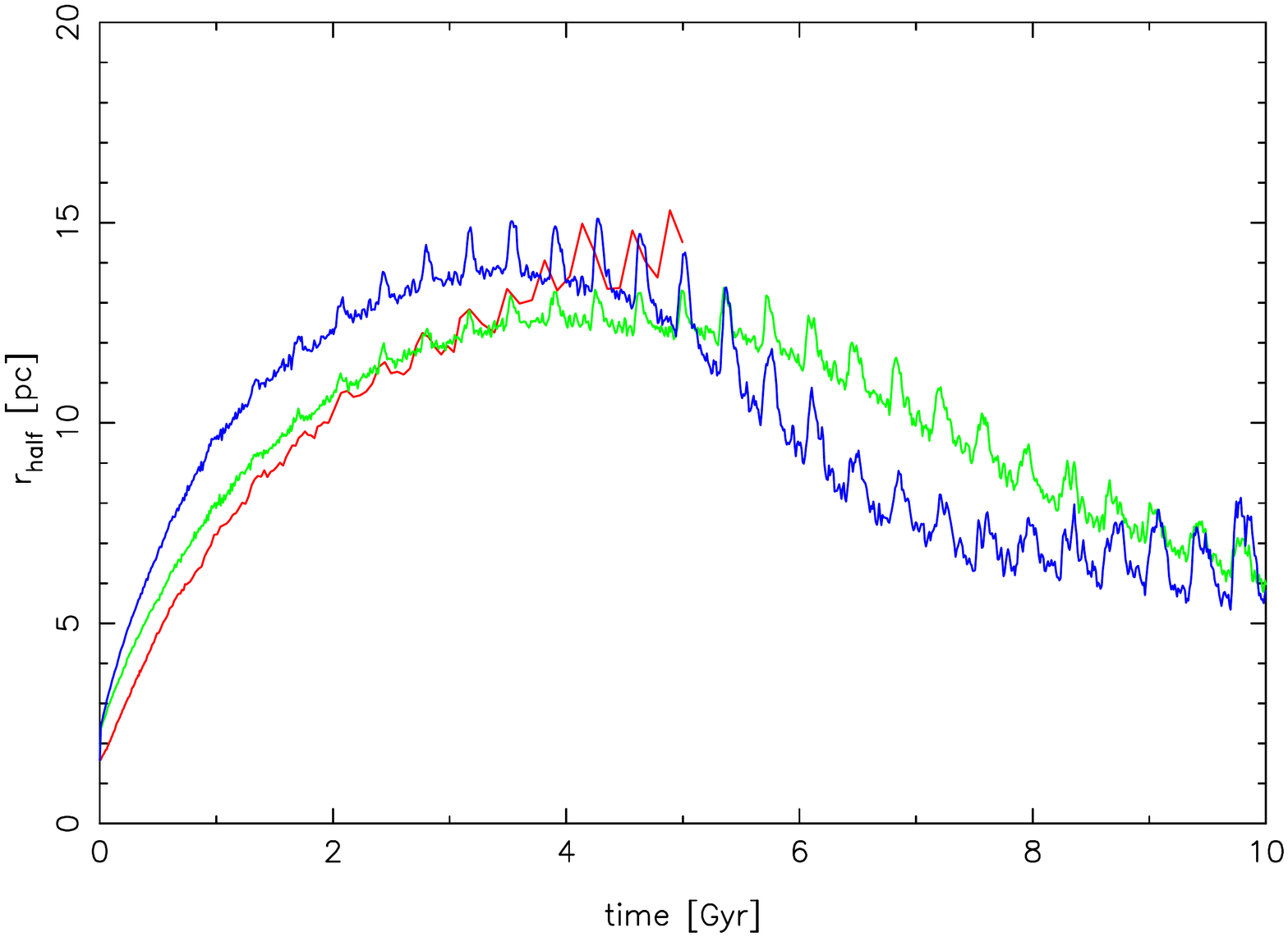}
\end{center}
\caption{A 20,000 Plummer sphere  cluster with mass $1.4 \times 10^4\msun$ created and evolved with
Nbody6 (red line) and the heated cluster code 
with the heating parameter set at 25 (the adopted value, blue) and 30 (green).
}
\label{fig_rhplum}
\end{figure}

The mass as a function of time of the clusters and their
half mass radii are displayed in
Figures~\ref{fig_mtplum} and \ref{fig_rhplum}, respectively, for the Nbody6 run and precisely 
the same cluster run with the heated cluster model used here. 
It is clear that with the right choice
of parameters the heated cluster model can reproduce a full n-body result to the
required accuracy. An important caveat is that one demonstration holds
for one cluster mass on one orbit.  
Some preliminary tests (Jeremy Webb private communication) have shown 
a degree of mass dependence of the most suitable heating parameters. The maximum radius inside of which
heating is applied to the star particles is fixed at $1.5r_v$, but the inner heating radius rises from 0.5 to 1.0$r_v$ 
for higher mass clusters.
The inner radius is described with the interpolating function,  
$r_{min}=0.5r_v(5.76\times 10^4\msun/M_c)^{1/3}$ over the mass range 
$M_c=5.76\times (10^3-10^4) \msun$, with the value pegged at the upper and bottom limit outside the range. 
The heating coefficient is similarly varied with cluster mass,
\begin{equation}
C_h= 36-9(5.76\times 10^4\msun/M_c)^{1/3}.
\label{eq_chm}
\end{equation}
A caution is that the heating coefficients are based on only two Nbody6 runs.
An important future step is to extend the cluster mass range, the range of galactic orbits, 
and the degree to which the reasonable range of stellar mass function, rotation, and binary stars effect the results.


\begin{figure}
\begin{center}
\includegraphics[angle=0,scale=0.8,trim=120 100 120 100,clip=true]{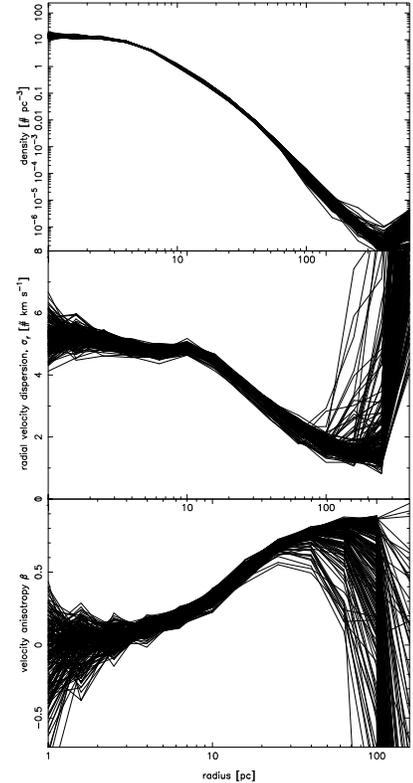}
\end{center}
\caption{The internal properties of the model clusters, density distribution (top panel), radial velocity dispersion (middle panel), 
and velocity anisotropy (bottom panel)  shown for all 198 clusters started as 
$W_0=7$ King models.}
\label{fig_GCkin}
\end{figure}

\subsection{Cluster Sizes and Halo Orbits}

\citet{King:66} models of star clusters are inserted into the 33 halos more massive 
than $0.5\times 10^{10}\msun$ in the z=3 start and  the 31 halos more massive than
$0.1\times 10^{10}\msun$ in the z=8 start.
Each star cluster starts with 20,000 star particles, with
particle masses of 15$\msun$ for clusters of total mass  $3\times 10^5 \msun$.  In
each halo, three clusters  are placed on near circular orbits at $R_m/2$ and another three $R_m/4$. 
The average $R_m$ is 8.4~kpc for the z=3 start and 2.6~kpc for the z=8 start, so the average cluster is at about 3 kpc
at z=3 and 1 kpc at z=8.
The clusters are started in a randomly oriented disk distribution,
at the local circular velocity with an added random velocity of 5\kms\ 
in each cardinal direction to minimize cluster-cluster interactions.

The King model star particle 
distribution is created with a potential parameter $W_0 = 7$, the ratio of the central
potential depth to the characteristic velocity dispersion of the cluster, resulting in a cluster with a ratio 
of tidal to core radius of 22.2.  A concentration
in this range is typical for the  \citet{Harris:96} [Fe/H]$\le -1$ clusters beyond 10 kpc galactocentric
radius.  The dark matter softening length is set at 2~kpc.  The star particle softening length is
set at 2 pc, comparable  to the size of a cluster core, which will not be resolved. However, the details of mass loss 
here are not significantly dependent on accurately following core properties. Reducing the star particle
softening to 1 pc sharpens the density profile somewhat, but makes essentially no difference to
the mass loss rate for an orbit in the logarithmic potential, at the cost of reduced step sizes.

A key consideration for the rate of cluster mass loss
 is the half-mass radius of the clusters. A King model has $r_h$ roughly one quarter of the outer radius of the cluster for
a wide range of central concentration.
If the model cluster of mass $M_c$ is scaled to the simple tidal radius, $r_t=r[M(<r)/M_c]^{1/3}$, 
where $r$ is the orbital radius and $M(<r)$ the halo mass within that radius.
For the sub-galactic halos the tidal radii work out to around
around 100 pc, which leads to $r_h$ of 25 pc, which is much larger than most current halo clusters. 
Evolving such large clusters in our n-body model leads to no surviving clusters
 within 100 kpc of the center of the final Milky Way-like halo. 
A somewhat artificial approach which avoids assumptions
is to scale the King model clusters to $r_t/10$ after which they naturally
 expand to their locally preferred size over the first Gyr of the simulation. 

The internal properties of all 198 clusters in the z=3 run are shown 1 Gyr
after the start in Figure~\ref{fig_GCkin}.  The core has expanded from an unresolved 0.5 pc to about 5 pc. 
The radial velocity dispersion shows the expected decline with radius then a rise into the unbound particles in the stream.
The cluster was started with an isotropic velocity dispersion, $\beta=0$, but with time  $\beta$ has become a rising
positive function, indicating a radially oriented velocity ellipsoid.

\begin{figure}
\begin{center}
\includegraphics[angle=0,scale=1,trim=140 180 140 180,clip=true]{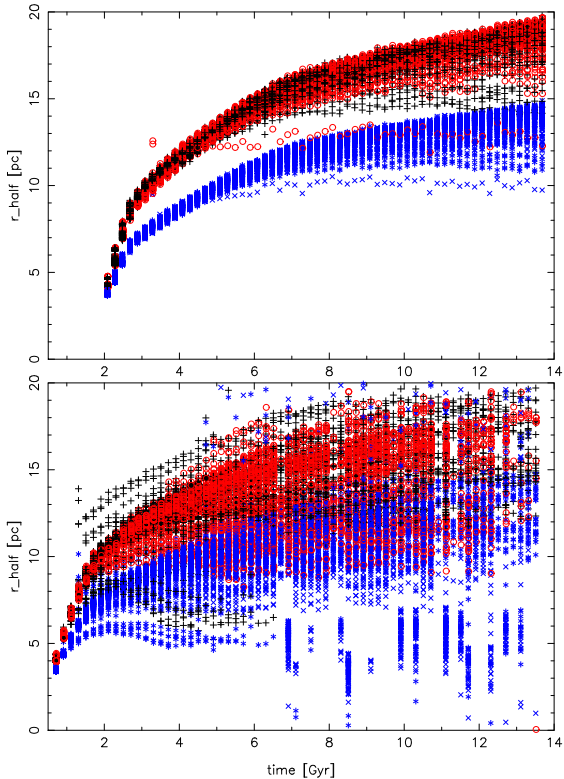}

\end{center}
\caption{The half mass (pluses) and virial radii (crosses), with clusters inside 100~kpc from the
main halo center shown as red or blue. 
The top panel is for the z=3 start and the z=8 start below. 
}
\label{fig_rhv}
\end{figure}

Figure~\ref{fig_rhv} shows the evolution of the cluster half mass radius and the virial radius within the clusters of the
z=3 (top) and z=8 (bottom panel) simulations. The z=8 clusters that fall into the main halo are completely evaporated over
the course of the simulation. The \citet{Harris:96} catalog gives a half light radius of 8 pc for low metallicity clusters
beyond 10 kpc galactocentric radius and 15 pc for those beyond 30 kpc. The half light radius is expected to be somewhat
 smaller than 
the half mass radius as a consequence of mass segregation in clusters. The clusters that remain at the end of the
simulations here typically have half mass radii of 15-20 pc. Overall the half mass radii in the simulations overlap
the range seen in Milky Way halo clusters, but the simulated clusters are missing
the very dense clusters with half mass radii of a few parsecs.  

\begin{figure}
\begin{center}
\includegraphics[angle=0,scale=1,trim=140 180 140 180,clip=true]{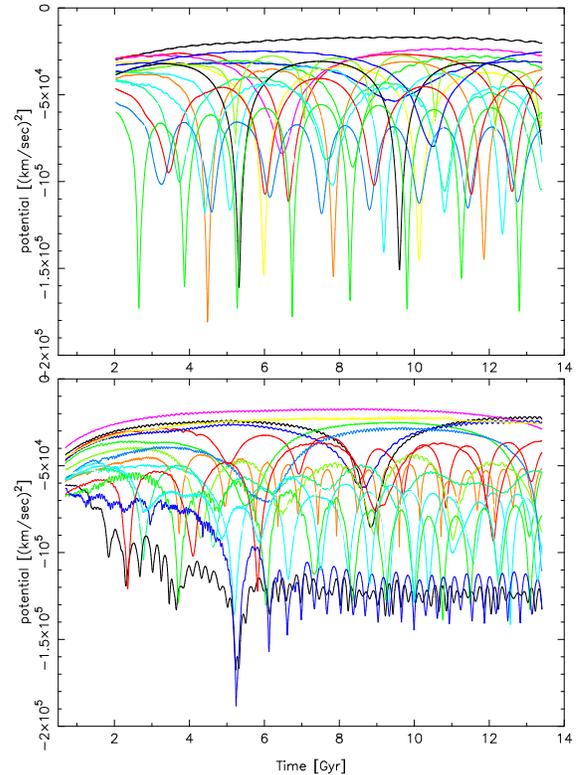}

\end{center}
\caption{The gravitational at the location of sixteen star clusters, one per halo. 
The top panel is for the z=3 start and the z=8 start below. The varying
potential depth is a combination of orbital motion within a sub-halo and the infall orbit 
into the growing galactic halo.
}
\label{fig_pot}
\end{figure}

\begin{figure}
\begin{center}
\includegraphics[angle=0,scale=1,trim=140 180 140 180,clip=true]{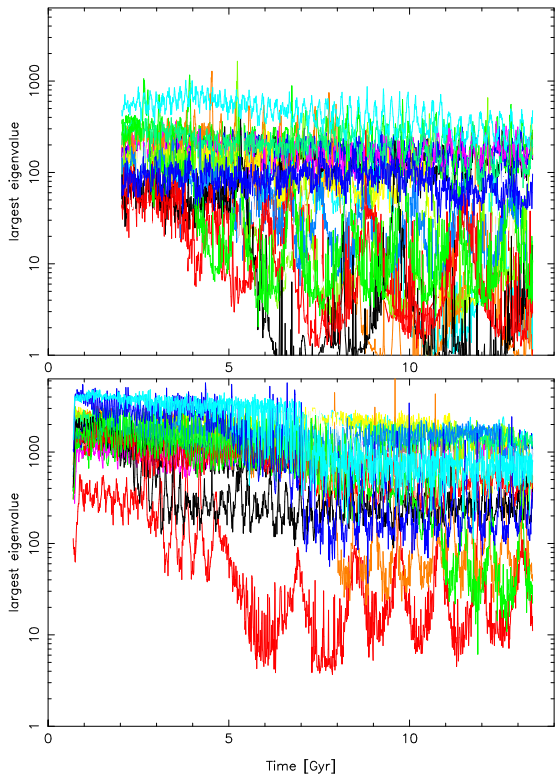}

\end{center}
\caption{The largest eigenvalue of the tidal tensor, in units of $(\kms/{\rm kpc})^2$, 
for one cluster per halo for sixteen halos.
The top panel is for the z=3 start and the z=8 start below. 
}
\label{fig_tides}
\end{figure}

The gravitational potentials at the location of a representative
 sixteen clusters in separate halos are
shown in Figure~\ref{fig_pot}. The z=3 simulation shows the pattern expected from accretion with little change
in radial period with time. The z=8 simulation shows a growing main halo that disrupts pre-galactic halos as internal orbital
motion ceases and as the radial oscillation period increases. The largest eigenvalue of the tidal tensors for the same
sixteen clusters is shown in Figure~\ref{fig_tides}. Tidal forces are dominated by the local density field which emphasizes
the variations in the tides due to local orbital motion in the pre-galactic halos, relative the motion in the main halo. The
tides are also somewhat noisy due to motion of nearby individual dark matter particles. The individual dark matter
particles here have a mass of $2\times 10^5\msun$ with a softening of 0.2~kpc, so individual particles 
will have a tidal field of about $100 (\kms/{\rm kpc})^2$ or less, with an encounter time of about 1~Myr, which is not resolved
in these plots at intervals of 10~Myr. The clusters are generally in regions where the dark matter particle 
density is of order $10^2$ particles per softening length.

\subsection{Numerical Considerations}

The dark matter particles are much heavier and faster moving than the star particles
and will slightly
heat the star particles over the course of the simulation. For encounters
at impact parameter $b$ and velocity $V$, over time $T$ the heating is the kinetic energy change per 
encounter, times the rate of encounters and the duration \citep{BT:08}.
The heating rate of the cool, light, star particles, per dynamical time, $t_d=V/R_s$, for a system of size $R_s$  with $N$ heavy
dark matter particles moving at velocities $V$, is then,
\begin{equation}
{d(\Delta V)^2\over dt} = {6 V^2\over N t_d} \ln{\Lambda} .
\label{eq_heat}
\end{equation}
where $\ln{\Lambda}$ is the Coulomb logarithm, with $\Lambda=R_s/s$,
with $R_s$ being about 100 kpc here, and $s$ is the dark matter particle softening. 
The VL-2 simulation had a force resolution of 0.162 kpc \citep{VL2}
of with particles of mass $1.8\times 10^5\msun$.
The simulation here has very similar particle properties, with
 dark matter particles of mass $2\times 10^5\msun$  and
a softening set to 0.2 kpc, to preserve the halo structure of the original simulation.  

The heating of the light, low velocity dispersion, star particles from heavy dark matter particles 
over a Hubble time needs to be below the velocity dispersion of the clusters, which 
is comparable to the spread of velocities in the tidal streams.  The simulation 
creates a halo with $\simeq 10^7$ halo particles with typical orbital velocities of 200\kms, 
for $\Lambda= 100/0.2$, $\ln{\Lambda}$ of 6, the heating over a Hubble time, about 20 dynamical times
in the inner regions of interest, is about 2\kms.  Lower mass cosmological halos have comparable dynamical
times, but fewer particles. Equation~\ref{eq_heat} indicates that heating varies as $V^{-1}$ for fixed $t_d$, 
so the heating will be about 6\kms in a 20\kms\ dark matter halo. 
The star clusters have characteristic velocity dispersions of 3 and 7\kms, at masses of $10^5\msun$ and $10^6\msun$, 
respectively. The heating is not negligible, but does not significantly compromise the results. 
The purpose here is to examine the basic effects. Large N simulations
will be done later.

The star clusters inserted into the simulation have an initial mass of $3\times 10^5\msun$, which is the 
mass of cluster 
at the peak of the current epoch globular cluster mass function. Each cluster in the primary simulations
begins with 20,000 particles, each of mass $15\msun$.  
The number of particles and the number 
of clusters inserted, about 200, are selected to give reasonable resolution of the tidal tails and a reasonable balance
between dark matter and star particles. Since the entire dark matter distribution needs to be evolved it makes sense to 
insert clusters in all the most massive pre-galactic sub-halos which comprise at least half of the dark matter mass.
Trial runs  found
that a star particle softening of 0.002~kpc allowed a reasonable time step, the shortest being about $2\times 10^4$ yr.
The star particle softening suppresses  relaxation giving the added heating of Equation~\ref{eq_dvheat}
complete control. 
Clusters of a fixed initial mass realized with ten times more particles were compared to these cluster
both run in a logarithmic external potential, finding essentially no dependence on particle number.

The Gadget2 parallelized tree code \citep{Gadget2} is very efficient for  cosmological simulations.
One complication in these simulations is the factor of $10^4$ scale difference between
the 100~kpc scale of the developing Milky Way-like halo and the star clusters with half mass
radii in the range of 10~pc. As the star clusters orbit through the dark matter background 
imbalances in 
memory and computational load between the processors develop
requiring relaxing various Gadget2 load balance parameters.
Runs  end at 13.4~Gyr, or 13.7 of the calculation's time units which is close to
the Hubble age for Via Lactea.
There are typically a total of approximately 200,000 time steps in
a simulation. 

\section{Cluster and Stream Evolution}

\begin{figure}
\begin{center}
\includegraphics[angle=0,scale=0.8,trim=180 180 180 180,clip=true]{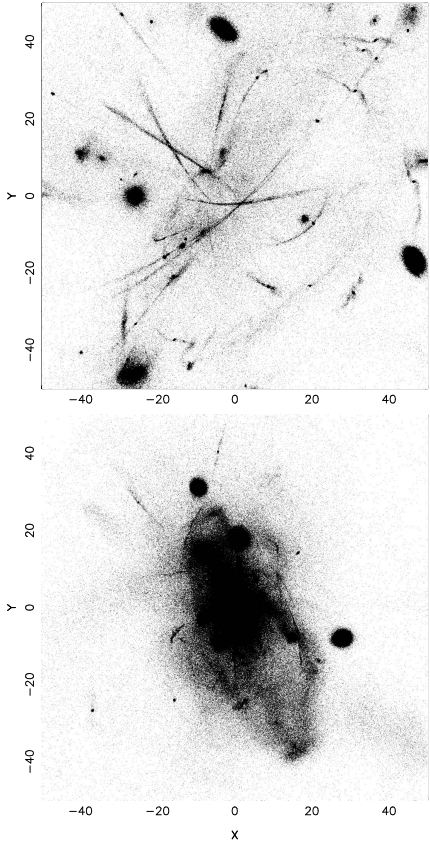}
\end{center}
\caption{The stars at the final moment of  the  z=3 (top) and z=8 (bottom) start simulations.
Only accreted systems are shown, with stars inserted in the dominant halo not plotted.
The boxes are 50 kpc center to edge or a 100 kpc on a side.
}
\label{fig_xysend}
\end{figure}

The distribution of the dark matter at the end of the
simulation is shown in Figure~\ref{fig_xydend} and the
stars at the same time are shown in Figure~\ref{fig_xysend} for the two runs.
The main dark matter halos in the two runs are comparable, with  masses of $9.3\times 10^{11}\msun$
and $7.5\times 10^{11}\msun$, for the z=3 and z=8 start simulations, respectively.
The z=3 start simulation has a more spherical halo.
Not visible in the frame is that both runs have 
a second large halo about half the mass of the dominant one, located
at a distance of 780 kpc for the z=8 start and 360 kpc for the z=3 start.

The two runs lead to very different cluster mass loss from the star clusters and spatial 
distribution of the tidal streams. The clusters in the z=3 run, 
top panel of Figure~\ref{fig_xysend}, have only a small 
 in a diffuse distribution and most of the star clusters remain intact. On
the other hand, the clusters in the z=8 run, bottom panel of Figure~\ref{fig_xysend}, have
 a substantial diffuse component and the clusters accreted on the main halo are completely dissolved.

\subsection{Tidal Streams and Remnant Clusters}

\begin{figure}
\begin{center}
\includegraphics[angle=-90,scale=0.85]{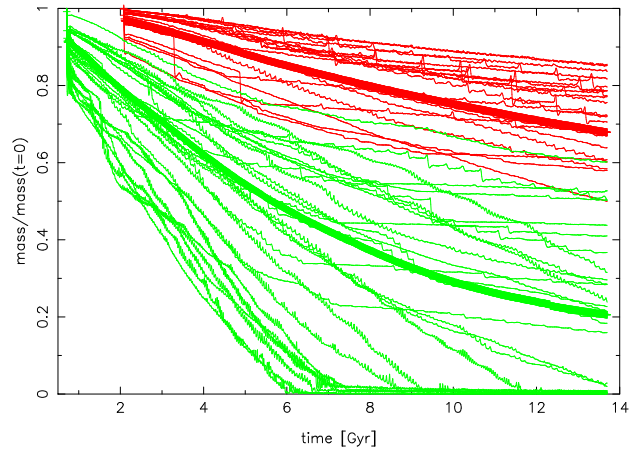}
\end{center}
\caption{Fractional mass remaining for a randomly selected 10\% subset of
the 198 (red) z=3 start and 186 (green) z=8 start clusters. All clusters
were started with masses of $3\times 10^5\msun$.
The heavy lines are the average fractional mass remaining of the clusters in the two simulations.
}
\label{fig_mt}
\end{figure}

The fractional  mass remaining in the clusters   as a function of time  is shown  in Figure~\ref{fig_mt}
for a random 10\% subset of the clusters in the two simulations.
Both simulations show a large range in cluster mass loss rates with time, 
reflecting the wide range of cluster tidal environments. The z=8 start clusters
are much more likely to be entirely disrupted before the simulation
reaches an age of half a Hubble time, about 7 Gyr.  In the z=8 start simulation,
larger halos build up as smaller halos encounter each other on nearly radial orbits,
so that their stellar contents pass through regions of strong tidal fields.
The z=3 start simulation has a large dominant halo into which significantly smaller halos fall
from relatively large distances and either miss or quickly pass through high tidal field regions, 
causing far less star cluster disruption.

\begin{figure}
\begin{center}
\includegraphics[angle=0,scale=0.8,trim=160 180 160 180,clip=true]{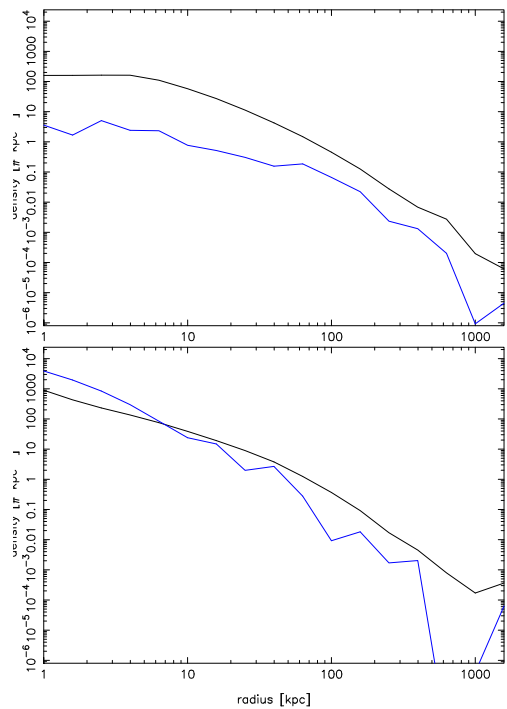}
\end{center}
\caption{Radial distribution of the
tidally removed stars (blue) and dark matter (black) for the z=3 (top) and z=8 starts (bottom panel).
}
\label{fig_dens}
\end{figure}

The radial distribution of dark matter (black lines) and stars stripped from the globular clusters (blue lines)
is shown in Figure~\ref{fig_dens} for the z=3 start (top panel) and z=8 start (bottom panel).
The star clusters are in relatively denser regions in the z=8 start than in the z=3 start, therefore relative
to the approximately the same overall dark matter distribution of the Milky Way-like final halo
the stars removed from the z=8 stars remain more bound and are a more centrally concentrated
population than the z=3 remnants. 


\begin{figure}
\begin{center}
\includegraphics[angle=0,scale=1,trim=170 240 170 240,clip=true]{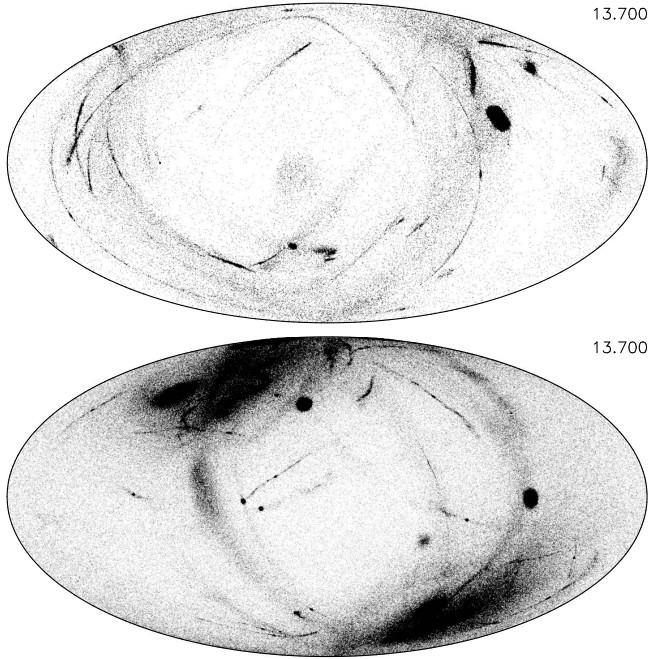}
\end{center}
\caption{The stars within 50 kpc (full weight with inverse distance squared weight on more distant star particles) of the galactic center 
at the end point for the  z=3 (top) and z=8 (bottom) runs
 projected on the sky. 
Only infall systems are shown, with clusters inserted in the dominant halo not plotted.
}
\label{fig_ah}
\end{figure}

The cluster stars are projected onto the sky  in Figure~\ref{fig_ah} using an equal
area Hammer-Aitoff projection. The center is defined as 
the center of mass of the clusters inserted into the main halo.
All stars to 50~kpc are plotted with equal  weights, with more distant ones reduced with the inverse
square of the distance to allow a somewhat realistic decline in visibility. As 
expected from the xy plots the z=3 start produces somewhat longer thin streams than the z=8 start.

Table~\ref{tbl-1} provides an overview of the basic statistics of the tidal streams in the two simulations. 
Streams are identified using the current observational
 technique, which is to identify relatively straight thin structures by eye. 
The minimum angular length that qualifies is approximately 10 degrees. 
These images have comparable numbers 
of visible particles (a few thousand typically) as real streams  have stars.
 The simulation has the advantage that there are effectively no 
field stars or dust that complicates finding Milky Way streams.

A few interesting conclusions can be taken from these two stream formation 
simulations, but a full analysis of stream properties is deferred to a future paper. 
In the z=3 simulation the infall clusters remain largely intact, with an average of 55\% of 
their initial mass retained. In complete contrast, the z=8 start has no remaining clusters in the inner regions
with more than 0.1\% of
their initial mass. 
The complete dissolution of the z=8 clusters may be at least partly due to 
fluctuations in the tides and will need to be tested in larger N runs.
The table shows that every z=3 start cluster inside 50~kpc has a visible stream
produces a visible stream, and 67\% of the clusters inside 100~kpc has
a visible stream. The z=8 start leads to a more compact distribution, 
with 78 cluster centers (no remnant clusters) inside 50 kpc,
but only 23 streams visible. Comparing the top and bottom panels of Fig~\ref{fig_ah} 
shows that the z=8 start streams are generally shorter than the z=3 start ones, 
likely a result of the z=3 infall being earlier and relatively smooth as indicated in
the potentials of Fig~\ref{fig_pot}. 
These two simulations effectively bracket the redshift
range over which globular clusters accreted onto our halo are formed. 
Because not all globular clusters have visible streams
the z=3 start is not a good explanation for all of the halo stream population, 
and, because no z=8 clusters survive in the main halo it also fails
to explain the full population, but the combination of the two, and intermediate times, 
would be a qualitatively acceptable mix.

The list of \citet{GC:16} finds that only 2 of the 12 streams have ``known or likely globular cluster progenitors", Pal~5 and  NGC~5466. 
Based on the result here that every accreted z=3  cluster should still
be present with its stream, one infers that 
10/12 of the streams have their origin in globular clusters formed
at redshifts greater than 3. Similarly the two streams with remnant clusters are likely to have formed around z=3 
and been accreted onto a dominant Milky Way-like halo fairly quickly. 
These results depend on the density of the region in which the clusters formed in their pre-galactic halos.
Clusters here were inserted at radii of the 1/4 to 1/2 of the peak of the circular velocity. 
If the clusters were inserted yet more deeply into the pre-galactic halos 
they would be in stronger tidal fields which would boost 
 the rate of dissolution upwards to that experienced in higher
redshift start simulations. 

The fact that  only 2 of the 12 thin streams have progenitor clusters. 
combined with the weak mass dependence of the rate of the evolution
of the cluster mass function seen here suggests that most globular clusters 
that were accreted onto the halo have dissolved.
Poisson statistics for 2 expected events indicates that up to 4 events contains 95\% of the probability, 
or, 2/3 of the cluster population that makes streams is dissolved. With the small sample and 
large extrapolation in mind, these numbers indicate that at high redshift that with
fairly high confidence there should be about 3 times more globular clusters that formed than
currently observed. 
More recent stream discoveries \citep{PS1,Grillmair:17}
find more streams and no more clearly associated globular clusters.

\begin{table}
\begin{center}
\begin{center}\caption{Cluster and Stream Counts.\label{tbl-1}}\end{center}

\begin{tabular}{lrrrrr}
\tableline\tableline
Start &  $M_c$ &$R_g$ max & $N_0$ & $N_f$  & Streams \\
\tableline
z=3 & $3\times 10^5\msun$ &50 kpc &  20 & 20 & 18 \\
      &  &100 kpc & 67 & 67 & 45 \\
z=8 &  $3\times 10^5\msun$&50 kpc & 78 & 0 & 28 \\
      & & 100 kpc & 87 & 0 & 29 \\
z=3 & $N \propto M^{-1.5}$ & 35 kpc & 50 &  45 & 28\\
\tableline
\end{tabular}
\end{center}
\tablecomments{Col 1: Starting time;
Col 2: $M_c$, cluster masses; Col 3: $R_g$ max, the maximum distance for full weight of the star particles, more
distant particles are de-weighted with the inverse distance squared;  Col 4: $N_0$, the number of cluster centers 
that fall into this volume; Col 5: $N_f$, number of clusters with at least 0.1\% of their initial mass remaining,
Col 6: the number of streams visible}
\end{table}

\begin{figure}
\begin{center}
\includegraphics[angle=0,scale=1.,trim=140 185 140 190,clip=true]{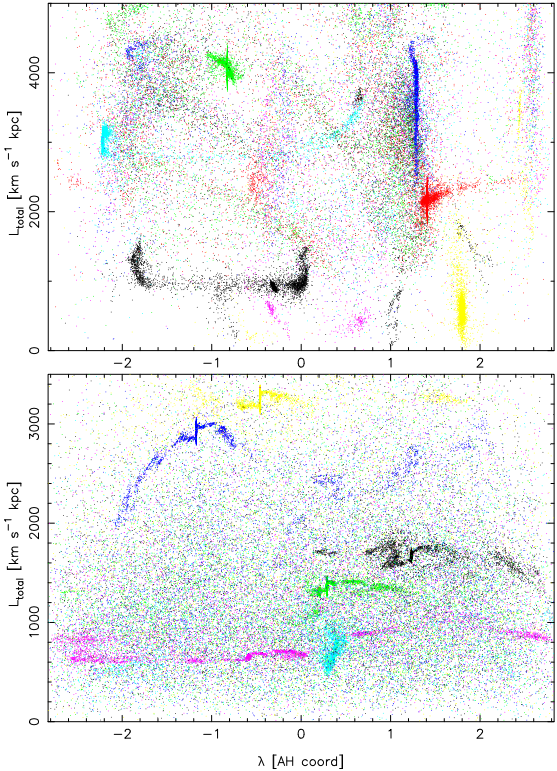}
\end{center}
\caption{The angular momentum calculated around the center of rotation 
in the inner galaxy for stars within 50~kpc of the
center. The top panel is for the z=3 start, the bottom for the z=8.
The points are color coded by their cluster of origin, cycling over 7 colors.
Approximately horizontal stripes are star streams. Near vertical stripes are
stripped stars orbitting within a sub-galactic halo.
The Aitoff-Hammer $\lambda$ variable has a range of $\pm2\sqrt{2}$ at the equator
but declines to zero at the pole.
}
\label{fig_lt}
\end{figure}

Although a dynamical analysis of streams is left to a future paper, streams are often modeled with 
their dynamical variables, such as angular momentum and energy. Figure~\ref{fig_lt} shows the
length of the angular momentum vector of individual star particles from a set of clusters that produce thin streams
in the two simulations. The horizontal variable is the same as in the Hammer-Aitoff plots of 
Figure~\ref{fig_ah}, which wraps around the sky. The angular momentum is calculated with
respect to the center of rotation of the inner-most star clusters. The halos here are triaxial, so the
angular momentum will not be a conserved quantity, but it remains sufficiently close to being conserved
that it is useful guide.

Thin tidal streams emanating from star clusters that are orbiting freely in the main
halo are the thin, approximately linear, features in Figure~\ref{fig_lt}. Stripped stars that
are orbiting within a sub-galactic halo create a vertical feature. A dark matter halo
that passes through a stream pulls stream stars towards the centerline of its path, leading
to a sideways ``S'-like feature in the angular momentum of the stream, which with time 
develops into a gap in the stream density. Both ``S'' and gap features appear to be present
in Figure~\ref{fig_lt}. 


\subsection{Star Cluster Mass Function Evolution}

\begin{figure}
\begin{center}
\includegraphics[angle=0,scale=1,trim=140 185 140 190,clip=true]{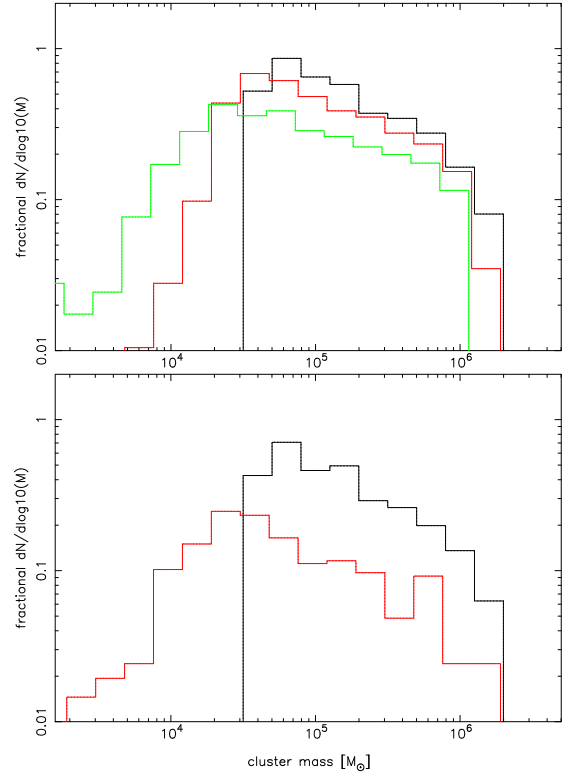}
\end{center}
\caption{Mass function of the globular cluster population  as 
a function of cosmic time for the z=3 (top) and z=8 starts (bottom panel).
In the top panel the times are 2.1 Gyr (black line), 7.9 Gyr (red), and 13.7 Gyr (green).
In the bottom panel the times are 0.7 Gyr (black) and 6.5 Gyr (red), with essentially no clusters
left at the current epoch. 
The initial population is proportional to $M^{-1.5}$ distribution between $5\times 10^4$ and 
$2\times 10^6 \msun$.
}
\label{fig_nm}
\end{figure}

\begin{figure}
\begin{center}
\includegraphics[angle=-90,scale=0.7, trim=100 70 100 70, clip=true]{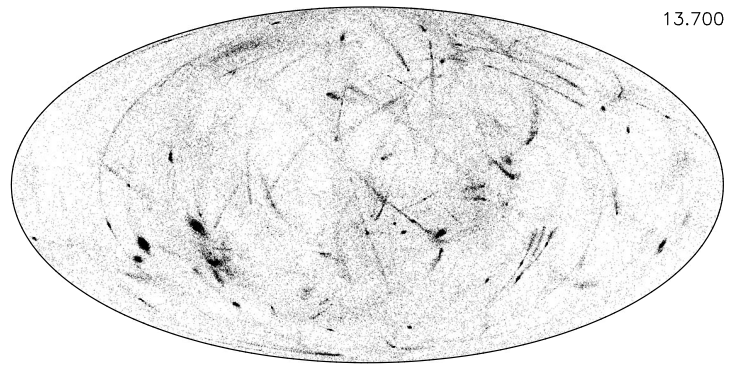}
\end{center}
\caption{The stars within 35 kpc at the end point for a z=3 run with a
$M^{-1.5}$ mass spectrum of globular clusters over the
range of $0.05-2.0\times 10^6 \msun$.
 projected on the sky. 
}
\label{fig_ahnm}
\end{figure}

The evolution of the mass function of the globular cluster population is of great interest for comparison to current
day globular cluster systems, and, for estimating the numbers of clusters that were present 
at high redshifts, within a few Gyr of formation.
To provide an initial estimate of the mass function
evolution  we populate the same dark matter simulations with
stellar clusters drawn from a $n(M)\propto M^{-1.5}$ mass distribution function, limited
to $0.05-2\times 10^6 \msun$. At total of 1433 clusters were inserted, distributed over all halos 
in proportion to the halo mass. The ratio of stellar mass to dark matter mass is $3\times 10^{-5}$ in the 
final main halo.
The mass functions are shown with time in Figure~\ref{fig_nm}.
The mean number over the inserted mass range declines to about 50\% of the initial population to a time of 13.4 Gyr.
On the other hand, the z=8 population is reduced by half at z=1 and disappears by redshift zero, again with
the caveat that this strong evolution needs further numerical verification.

An interesting outcome is that the mass loss from the clusters has a relatively weak rise towards lower masses. 
The significantly more rapid destruction of clusters in the z=8 start as compared to the z=3 start can
be attributed to the significantly stronger tidal forces that the clusters in the high redshift start experience. Although the
z=3 start clusters have a lower rate of mass loss, they also show little mass dependence, suggestive that
tidal heating continues to overwhelm the internal two-body heating. 
An important caveat is that these results depend on both the internal heating model used,
 and, the fluctuations in the tidal field that the current particle numbers allow.

The sky distribution of the variable mass clusters is shown in Figure~\ref{fig_ahnm}. Here the maximum distance
for full weight visibility is 35 kpc, to be comparable to the \citep{PS1} maps. It is important to note that the most
massive clusters make the most visible streams, because their lose stars at a higher absolute rate.  
\citet{PS1} find 15 streams in $3\pi$ of sky to a depth of 35 kpc, although about another half dozen streams are known
from other data. The plane of the Milky Way reduces the effective sky area for stream finding significantly. If it is reduced
to $2\pi$, or half the total sky, then the  28 streams in the simulation,Table~\ref{tbl-1}, is
in reasonable agreement with the streams in the sky, although all of these have remnant clusters in them, which the
real sky does not.

Globular clusters are a site for the creation  of some fraction of the
 relatively heavy binary black holes that LIGO has detected \citep{LIGO1,LIGO2} through
the gravitational hardening and exchanges of binaries in the cores of globular clusters \citep{PZM:00,Oleary:16,
Mapelli:16,Hurley:16,Rodriguez:15,Rodriguez:16,Askar:17,Park:17}. The binary black holes
that are merging today formed at much earlier times. 
The simulations here indicate that the clusters above $10^5\msun$  that have survived are 1/2 or less of
the high redshift globular cluster population when the BBHs were produced.
The prediction of the cluster binary  LIGO event rate 
is therefore at least twice times the event rate that  is predicted based on the numbers of globular clusters, with
the statistics indicating that in the mean a full factor of five more could have been present.
More recent stream discoveries \citep{PS1,Grillmair:17} pushes the number of streams up another 5-6 streams,
with one somewhat marginal case for a progenitor globular cluster \citep{Koposov:14,PS1}.
At present, these numbers rest on 12-18 streams, 2-3 globular clusters, and two simulations. Both
additional simulations and a more quantitative comparison to sky data are needed to harden these conclusions.

\section{Discussion and Conclusions}

Stellar dynamical, globular cluster-like, star clusters are placed in a 
 in a full cosmological simulation at approximately redshifts 3 and 8.
The simulations give a cosmologically realistic view of how the thin, and not-so thin,
stellar streams from globular clusters are formed and the overall spatial
distribution of the stars evaporated and tidally pulled from globular clusters. 
The simulations are gravity-only, with no hydrodynamics or star formation mechanism.
The Gadget2 code is augmented to introduce an appropriate level of two-body driven heating
into the star clusters, with heating parameters calibrated with Nbody6 \citep{Aarseth:99} runs of a few clusters.
The current calibration runs are limited to low mass clusters and one or two orbits. The modified Gadget2 code
is able to reproduce the calibration runs to within about 10\% in mass loss and cluster size. These initial runs 
typically have $10^7$ dark matter particles and a comparable but generally somewhat smaller number
of star particles. The particle numbers are sufficient, but just barely, to keep heating of star particles from
dark matter particles below the very low velocity dispersion present in the streams.  The resulting star clusters
at the end of the simulation have half mass radii in the range of those seen in the halo, 15-20 pc, but 
tend to be somewhat larger.  These and other numerical issues will be examined further with additional simulations.

The basic stream morphologies are comparable to observed streams, with the simulations giving thin streams lengths 
ranging from around 10 degrees up to about 100 degrees on the sky. Essentially all streams have an 
extended diffuse component that is several kpc wide that is composed of stars tidally removed from the clusters 
while they were in their pre-galactic halos which upon accretion into the growing Milky Way halo 
are spread out over the width of their orbit in the dwarf.  The longer thin streams generally exhibit 
density variations along their length, 
qualitatively as expected on the basis of dark matter sub-halos crossing the streams and creating gaps.

The survival rate of the z=3 start clusters with streams is 100\% inside 50kpc, whereas the z=8 start has no surviving
clusters in the inner halo. The observed streams have 2 or 3 visible progenitor clusters in 12-18 streams \citep{GC:16,PS1}.  
Although the statistics have large uncertainties a tentative
conclusion is that for every massive current epoch globular cluster there were at least two, and possibly more, at high redshift.  

The mass loss is, appropriately, sensitive to the details of the heating model for the star clusters. The model has been validated for only a few orbits 
with low mass clusters. In particular the significant tidally driven evolution of 
the massive clusters can be understood as tidally driven, but will be examined further in future n-body simulations.


\acknowledgements

Comments from an anonymous referee spurred substantial revisions to this paper.
This research was supported by  NSERC of Canada.

\end{document}